\documentclass{appolb}
\usepackage{graphicx}

\usepackage{amsmath,amssymb}
\usepackage{CJKutf8}
\usepackage{dcolumn}
\usepackage{grffile}
\usepackage{tikz}
\usepackage{bm}
\usepackage{slashed}
\usepackage{physics}
\usepackage[colorlinks=true,linkcolor=blue,citecolor=blue, urlcolor=blue]{hyperref}
\usepackage{xcolor}

\usepackage{asymptote}

\def\6{\partial}

\newcommand{\be}{\begin{equation}}
\newcommand{\ee}{\end{equation}}
\newcommand{\bea}{\begin{eqnarray}}
\newcommand{\eea}{\end{eqnarray}}

\begin{document}
\title{Opportunities with ultra-soft photons: Bremsstrahlung from stopping%
\thanks{Presented at Quark Matter 2022}%
}
\author{Sohyun Park 
\address{Theoretical Physics Department, CERN, CH-1211 Gen\`eve 23, Switzerland}
}
\maketitle
\begin{abstract} 

We compute the spectra of bremsstrahlung photons for different stopping scenarios which give rise to different initial charge-rapidity distributions. In the light of novel experimental opportunities that may arise with a new heavy-ion detector ALICE-3 at the CERN LHC, we discuss how to discriminate between these stopping scenarios and how to disentangle bremsstrahlung photons from other photon sources. 
 \end{abstract}

In the initial stage of ultra-relativistic nucleus-nucleus collisions, the electric charges of the incoming nuclei are decelerated significantly. This charge deceleration induces bremsstrahlung whose intensity is sensitive to the degree of longitudinal stopping. This has led to the idea of using bremsstrahlung from stopping to constrain the longitudinal charge distribution right after the collision~\cite{Kapusta:1977zb,Bjorken:1984sp,Dumitru:1993ph}. 
Refined calculations of classical electromagnetic bremsstrahlung were carried out already in the late 1990s and they demonstrated that the effects are measurable~\cite{Jeon:1998tq,Kapusta:1999hb,Wong:2000hka}. In this regard, a dedicated detector at the BNL Relativistic Heavy Ion Collider (RHIC) was proposed~\cite{Jeon:1998tq} but it was never realized. 

With our recent work~\cite{Park:2021ljg} and with the additional calculations reported here, we aim to reassess the time-honored discussion of ~\cite{Kapusta:1977zb,Bjorken:1984sp,Dumitru:1993ph,Jeon:1998tq,Kapusta:1999hb,Wong:2000hka} for the different kinematic range at LHC and in light of the new experimental opportunities that the new ALICE-3 detector may offer in coming decade~\cite{Adamova:2019vkf}. In particular, this detector may give experimental access to ultra-soft photons with transverse momentum as low as 10 MeV at forward rapidity up to $y = 4$. As explained below, this is a potentially interesting kinematic range for the study of bremsstrahlung photons.

\section{Classical bremsstrahlung}
For an electric current $\mathbf{J}(\mathbf{x}, t)$ that varies in space and time, the intensity of photons emitted with energy $\omega$ in direction $\mathbf{n}$ is given by 
the classical bremsstrahlung formula~\cite{Jackson:1998nia}, 
\begin{equation}
	\frac{d^{2} I}{d \omega d \Omega}=|\mathbf{A}|^{2}\,,  ~\mathbf{A}(\mathbf{n}, \omega)=\int d t \int d^{3} x \mathbf{n} \times(\mathbf{n} \times \mathbf{J}(\mathbf{x}, t)) e^{i \omega(t-\mathbf{n} \cdot \mathbf{x})} \,.
	\label{eq:Jackson_14.67}
\end{equation}
In heavy-ion collision, up to the collision time $t=0$, the electric charges move on the Lorentz-contracted pancakes of the incoming nuclei. We describe them by 
incoming currents $J^{(in)}_{\pm}({\bf x},t)$, written in terms of the incoming charge density $Z\, e\, \rho_{\rm in}(r)$ in the transverse plane to the beam direction $z$ times the beam velocity $v_0$,
\begin{equation}
    J^{(in)}_{\pm}({\bf x},t) = \pm \, \Theta(-t) \,Z\, e \, \rho_{\rm in}(r) \, v_0\, \delta(z \mp v_0t)\, .
    \label{eq1}
\end{equation} 
Here, ${\bf x} = ({\bf r},z)$, and we write the incoming velocity   $v_0 = \tanh y_0$ in terms of the projectile rapidity $y_0 \simeq \ln\left(\frac{\sqrt{s_{\rm NN}}}{m_N}\right).$ 
The entire electrical current $\mathbf{J}$ is the sum of the incoming ($t < 0$) and outgoing ($t>0$) pieces, 
$	\mathbf{J} = \mathbf{J}^{(in)}_{+} + \mathbf{J}^{(in)}_{-} + \mathbf{J}^{(out)} $. 
The outgoing piece $\mathbf{J}^{(out)}$ depends on the dynamics of stopping, \emph{i.e.}, on the
charge-rapidity distribution $\rho({\bf r},y,t)$  right after the collision, 
\begin{equation}
    J^{(out)}({\bf x},t) = \Theta(t)\, \int_{-y_0}^{y_0} 
    \rho({\bf r},y)\, 
    v(y)\, \delta\left(z - v(y)t \right)\, dy\, ,
    \label{eq2}
\end{equation} 
where $-v_0 < v(y) = \tanh y< v_0$.
Since the deceleration is primarily in the longitudinal direction, we consider in the following electric-charge density distributions in which the longitudinal and transverse dependence
factorizes, 
$	\rho({\bf r},y) = \rho_{\rm in}({\bf r})\, \rho(y) $. 
The intensity of photons in Eq.~(\ref{eq:Jackson_14.67}) then can be written as 
\begin{eqnarray}
\frac{d^2I}{d\omega d\Omega} = \frac{\alpha Z^2}{4\pi^2}\sin^2\theta \left| F(\omega R \sin\theta) \right|^2 \!
\left| \!
\left[ \int \! dy \frac{v(y) \rho(y)}{1\!-\!v(y)\cos\theta} \!-\! \frac{2v_0^2 \cos\theta}{1\!-\!v_0^2\cos^2\theta} \right] \!\right| ^2 \,.
\label{eq8}
\end{eqnarray} 
Here the transverse form factor $F$ is the Fourier transformation of charge density in the transverse plane,
\begin{equation}
F(\omega R \sin\theta)=
\int d^2r_{\perp} \, \rho_{\rm in}\left( r_{\perp} \right)\, e^{-i \omega {\bf n}\cdot {\bf r}_{\perp}} \, ,
\end{equation} 
which can be approximated by assuming that charges are distributed homogeneously in a sphere of radius $R$,
\begin{equation}
F(q) = \frac{3}{q^2}\left( \frac{\sin q}{q} - \cos q \right)\,, ~ q \equiv \omega R \sin\theta \,.
\end{equation} 
This form factor becomes unity for a sufficiently small $q$. 
This means that the intensity 
is insensitive to the transverse profile of the charge distribution for sufficiently soft (small $\omega$) and collinear (small $\theta$) photons, \textit{i.e.}, $q = \omega R \sin\theta \ll 1$. 
Indeed the classical formula is valid for soft and collinear photons that do not resolve internal structure~\cite{Park:2021ljg} .
Hence, we work exclusively in the kinematic regime of $\omega < 200 \mbox{ MeV},\, \eta > 3$ which ensures $q <1$.  

\section{Modelling the charge-rapidity distribution}

\begin{figure}[htb]
\centerline{%
\includegraphics[width=12.5cm]{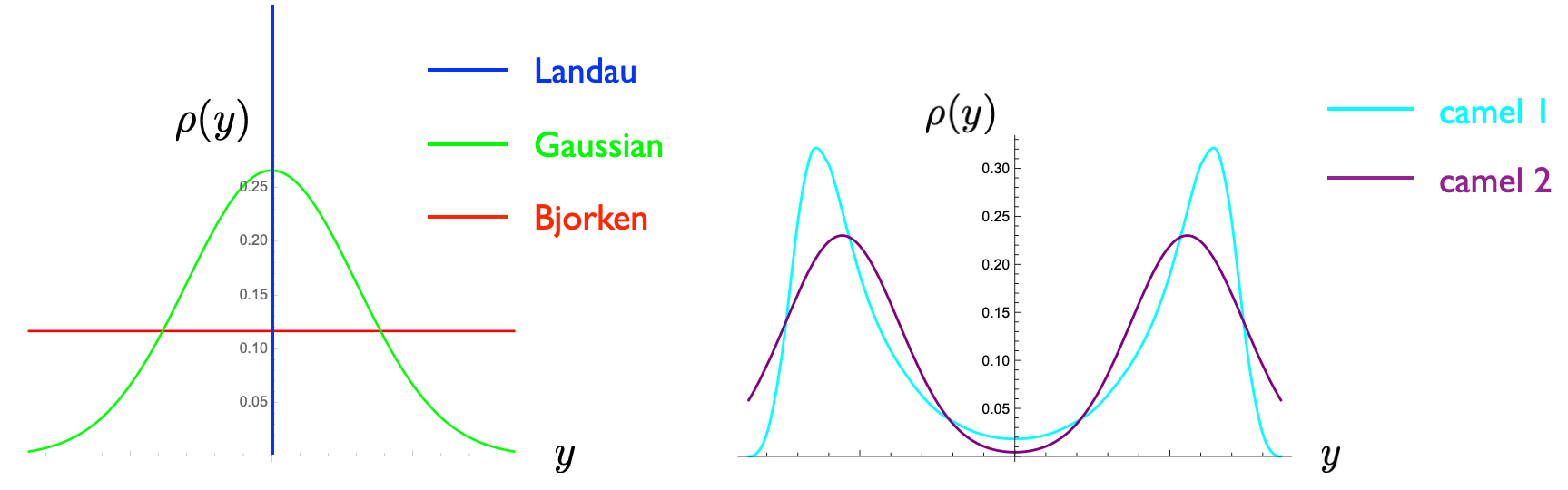}}
\caption{Five different models of charge-rapidity distributions of net charge in a nucleus-nucleus collision, referred to as 
Landau (in blue), Gaussian (in green),    Bjorken (in red), camel 2 (in purple) and  camel 1 (in cyan).
}
\label{fig:charge-rapidity-distributions}
\end{figure}

The only unknown in the above set-up  is the charge-rapidity distribution $\rho(y)$ in Eq.~\eqref{eq8}. The shape of $\rho(y)$ determines the bremsstrahlung spectrum and measuring this bremsstrahlung  (or not measuring it but establishing a tight upper bound on it) constrains the allowed stopping scenarios $\rho(y)$. Here, we conmpare the five different 
models of charge-rapidity distribution depicted in Fig.~\ref{fig:charge-rapidity-distributions}.  The total charge in the final state is $2 \times Z$ where $Z$ is the charge number of the incoming nucleus. We adopt in the following a normalization where the integral over $\rho(y)$ is normalized to 2.

\section{Numerical results}

\begin{figure}[htb]
\begin{center}
  \begin{tabular}{cc} 
 \includegraphics[width=0.45\textwidth]{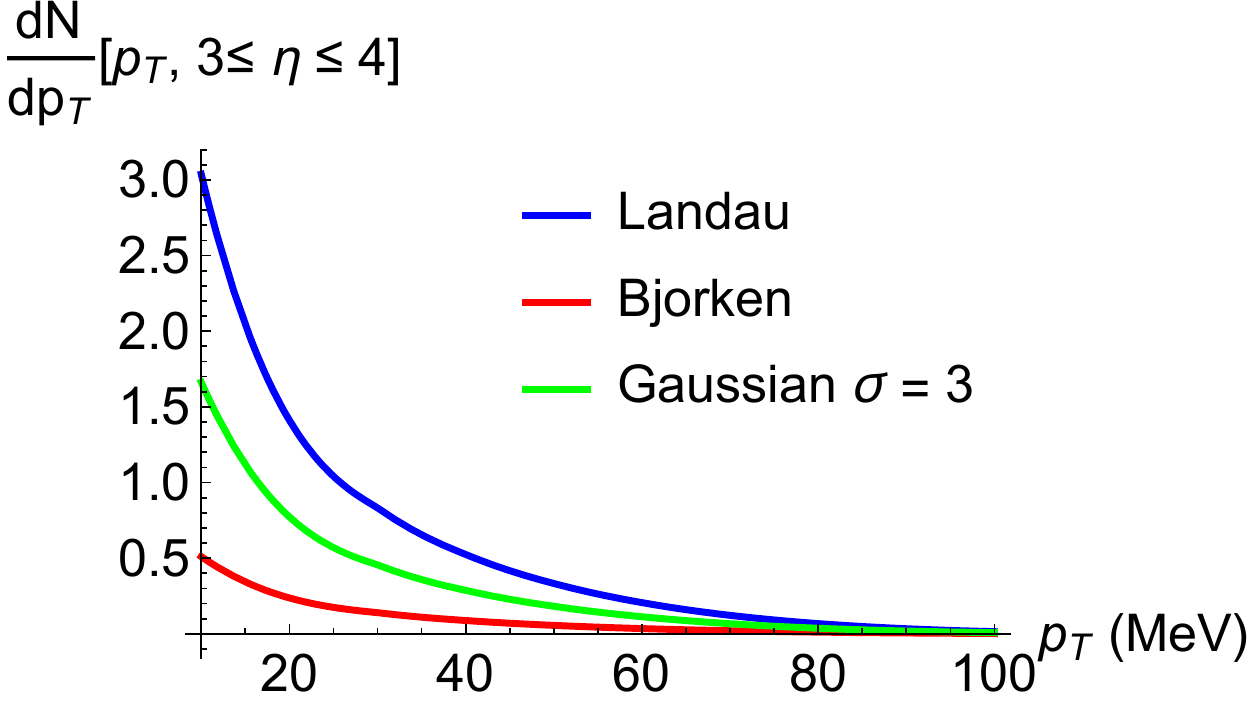}  
    &
  \includegraphics[width=0.45\textwidth]{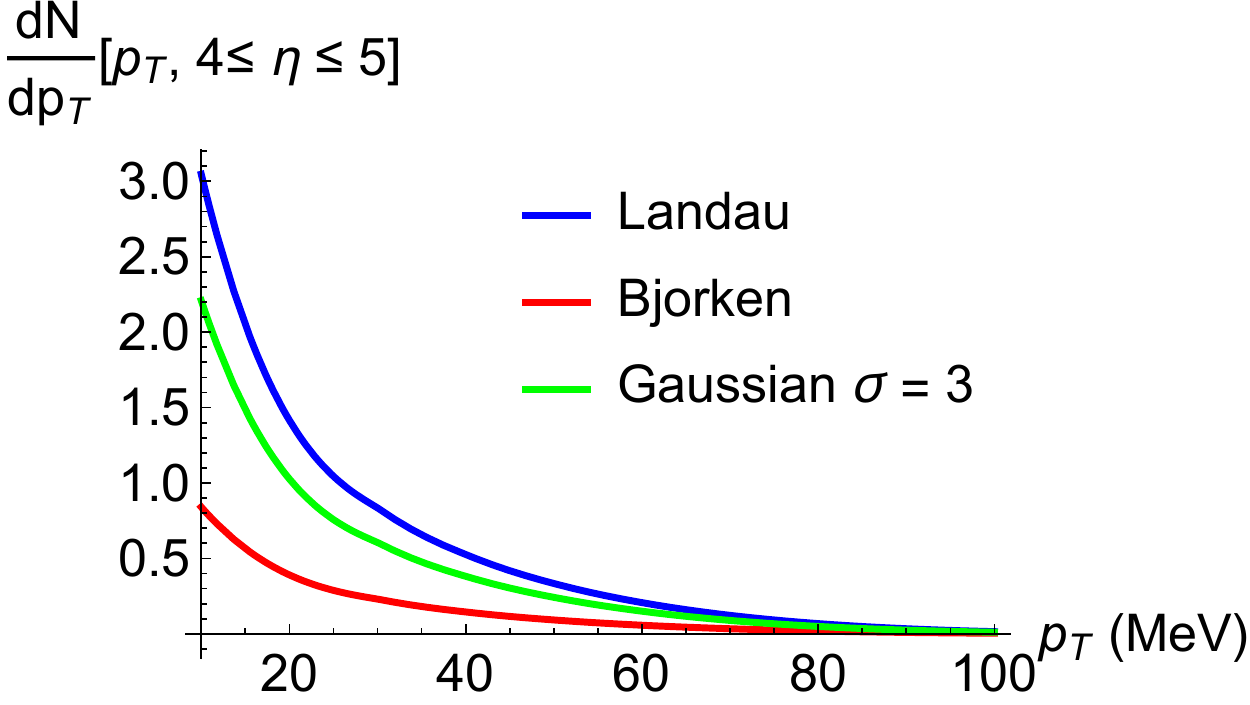}  
  \end{tabular}
 \begin{tabular}{cc}
 \includegraphics[width=0.45\textwidth]{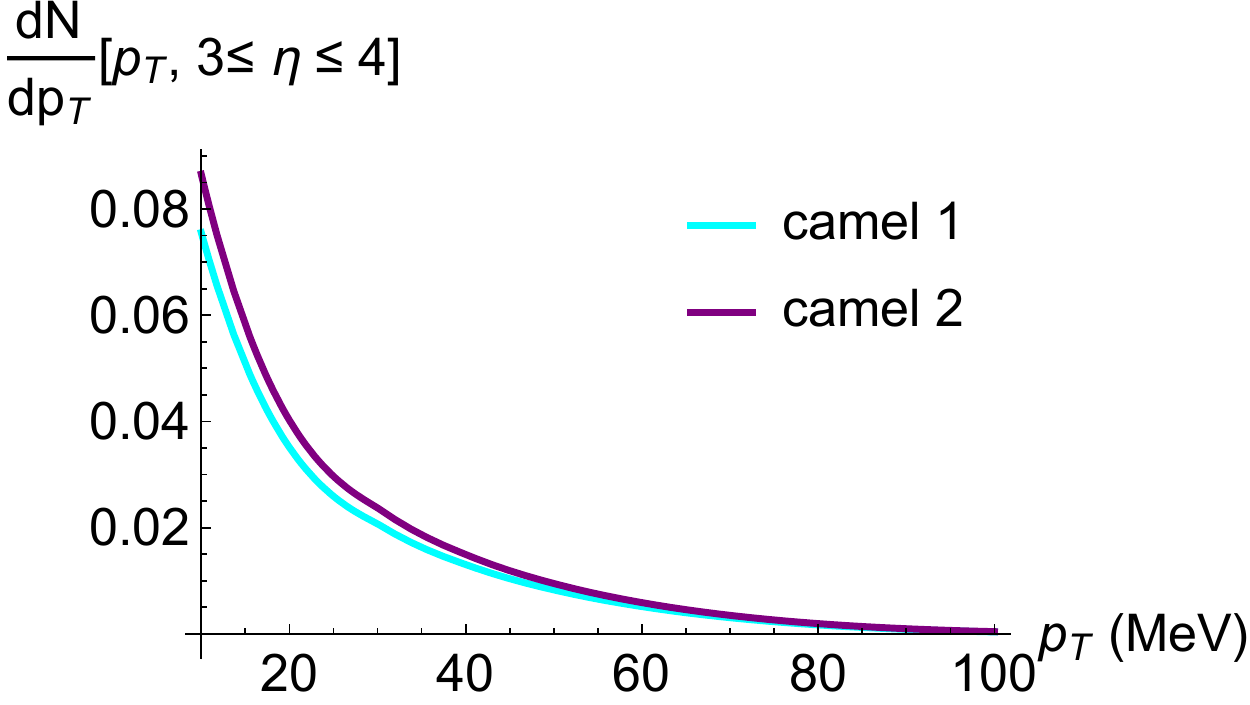}  
    &
  \includegraphics[width=0.45\textwidth]{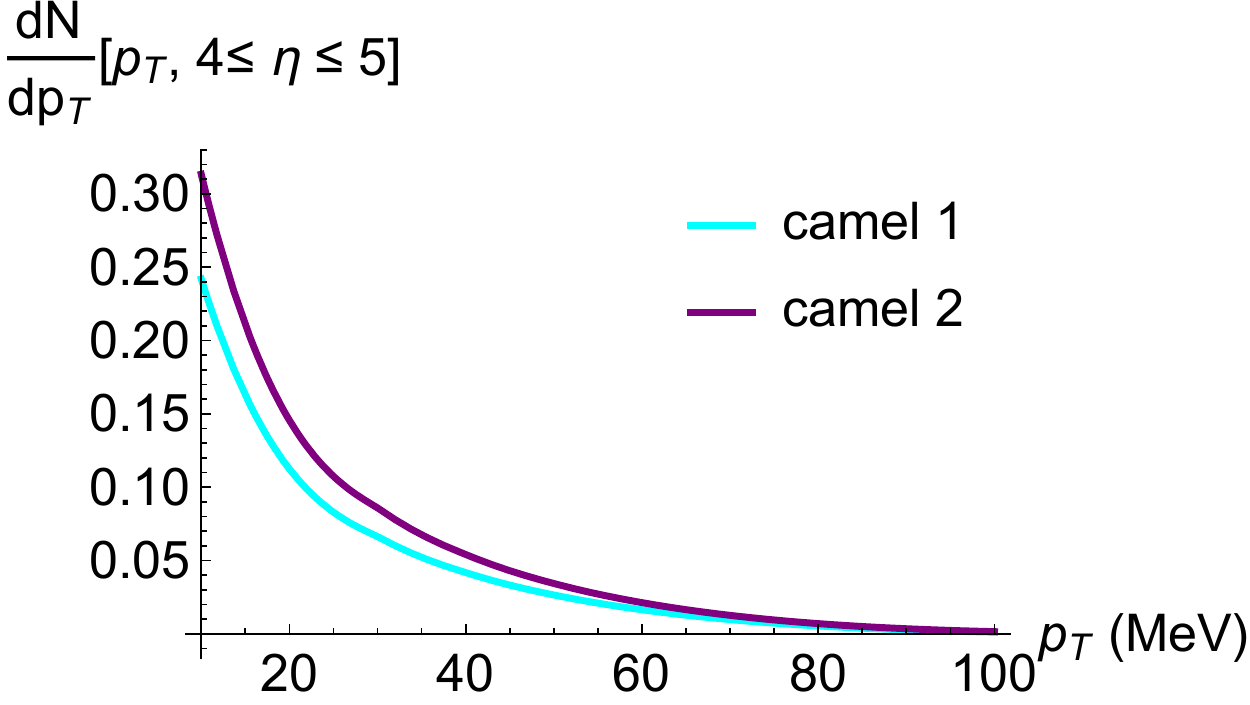}  
  \end{tabular}
\end{center}
\caption{The differential photon number spectrum as a function of $p_T$ for different $\eta$ bins.} 
\label{fig:differential-photon-number}
\end{figure}

\begin{figure}[htb]
\begin{center}
  \begin{tabular}{cc} 
 \includegraphics[width=0.45\textwidth]{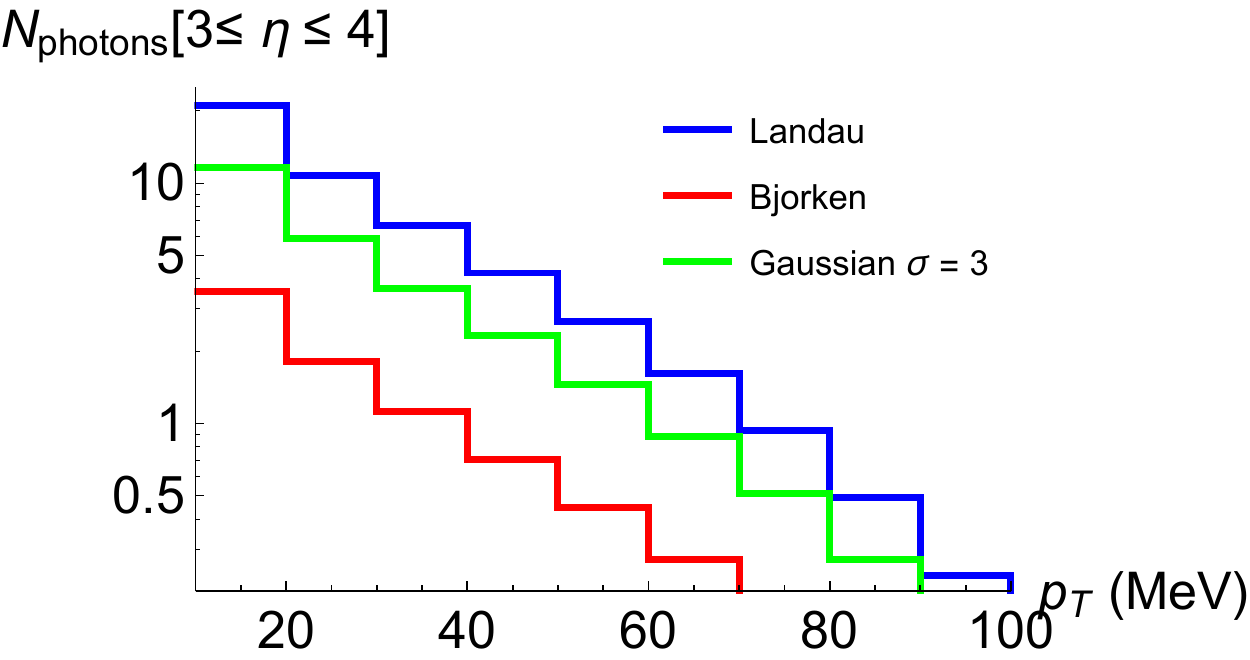}  
    &
  \includegraphics[width=0.45\textwidth]{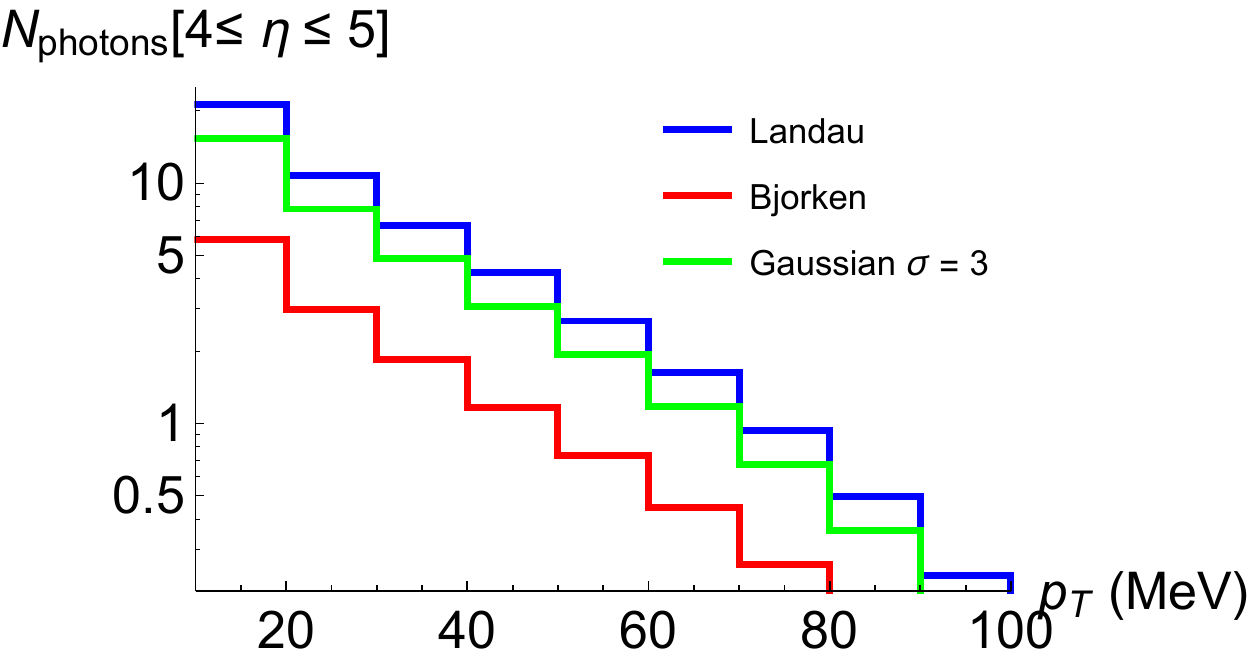}  
  \end{tabular}
 \begin{tabular}{cc}
 \includegraphics[width=0.45\textwidth]{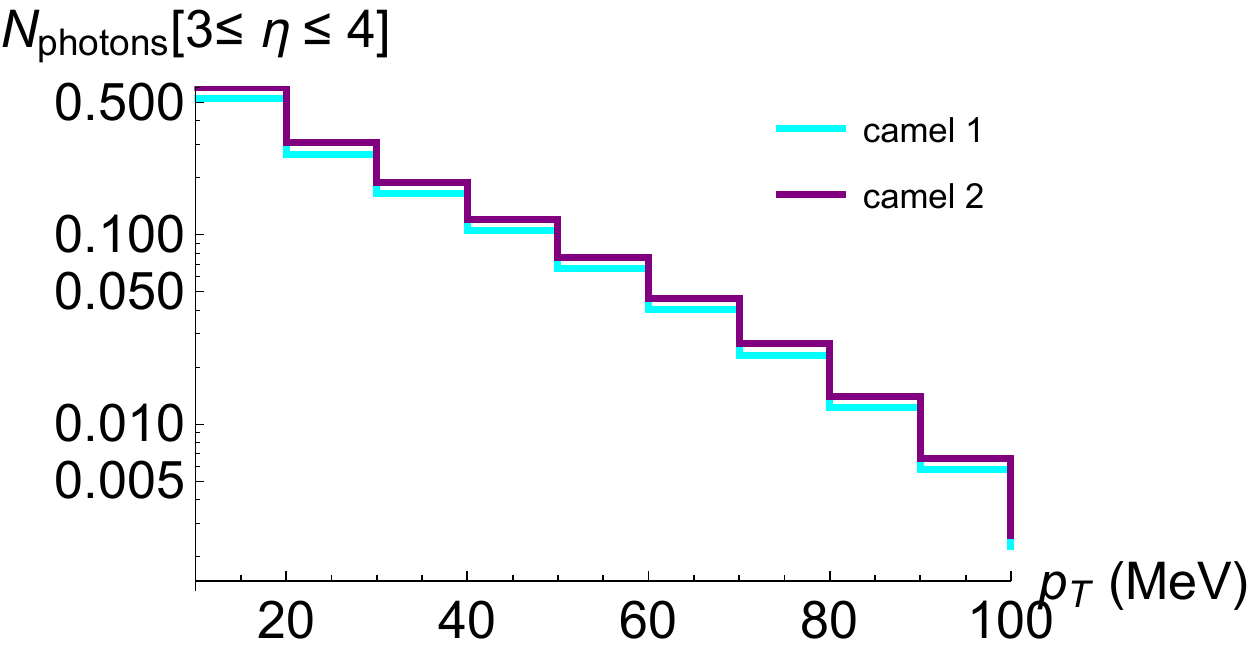}  
    &
  \includegraphics[width=0.45\textwidth]{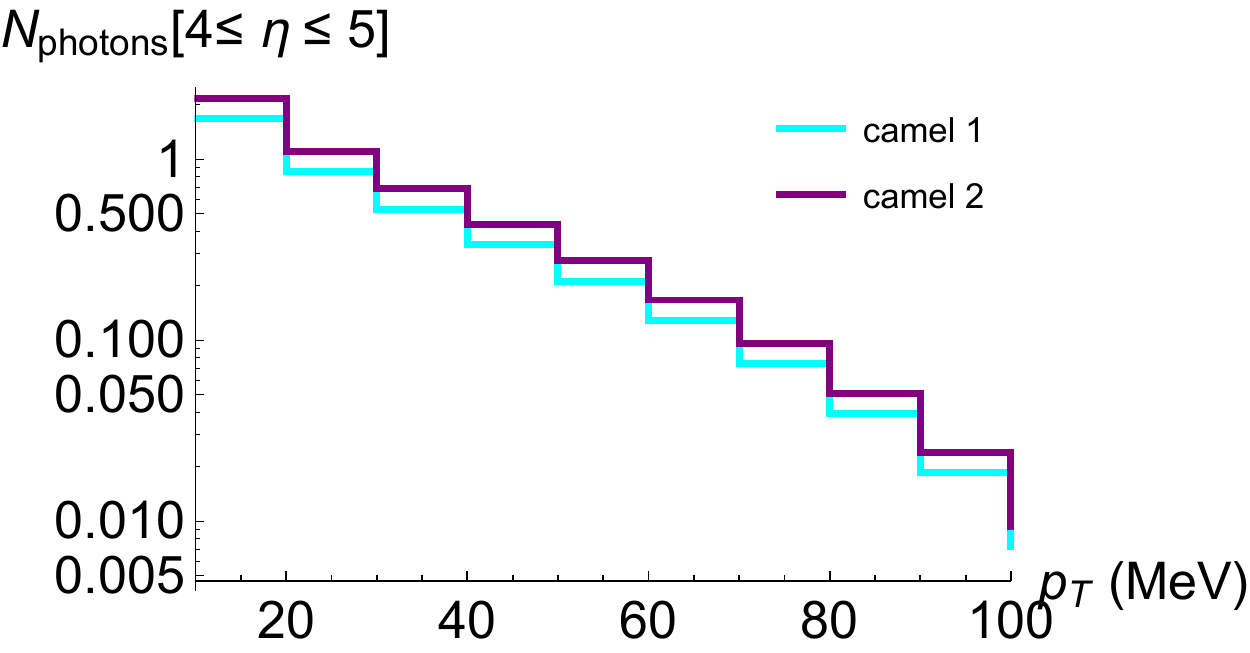}  
  \end{tabular}
\end{center}
\caption{The number of photons per unit rapidity window in $p_T$ bins of 10 MeV} 
\label{fig:photon-number}
\end{figure}

Changing the kinematic variable in \eqref{eq:Jackson_14.67} from photon energy and angle $(\omega, \theta)$ to transverse momentum and pseudo-rapidity $(p_T, \eta)$, we determine the the number of photons radiated per unit phase space:
\be
	\frac{d^2N}{dp_T d\eta} = \frac{2\pi}{p_T\, \cosh^4\eta} \frac{d^2I}{d\omega d\Omega}\, .	
	\label{eq18}
\ee
Fig.~\ref{fig:differential-photon-number}  shows the resulting $p_T$-differential bremsstrahlung spectrum with the characteristic soft  $1/p_T$ divergence 
and Fig.~\ref{fig:photon-number} shows the resulting number of photons per unit rapidity window in $p_T$ bins of 10 MeV.

The five stopping scenarios studied here map out different extremes: in the Landau scenario, all charges are maximally stopped and this scenario thus yields the largest photon bremsstrahlung intensity. The other four models (Gaussian, Bjorken, Camel 2 and Camel 1) correspond to scenarios in which less and less stopping occurs. In the limit in which $\rho(y)$ would correspond to two Kronecker delta's at rapidities $\pm y_0$, no stopping would occur and therefore, no bremsstrahlung would be radiated. The model ``Camel 1'' comes the closest to this limiting case and it is therefore accompanied by the smallest photon bremsstrahlung intensity. Our numerical result quantifies this dependence.

\section{Background photons}
Photons from $\pi^0$-decays are expected to be the dominant background. For a rough estimate, we note that the $p_T$-differential charged pion spectrum is almost flat for soft 
$p_T$ at mid-rapidity:
$dN^{\pi^\pm}/dp_T\, dy \simeq 2000/{\rm GeV}$ for $p_T < 500$ MeV at $y=0$ (read off from Fig. 21 of \cite{Nijs:2020roc}
which replots data from ~\cite{ALICE:2014juv}).
The $\eta$-differential charged pion spectrum $dN_{\rm ch}/d\eta$ decreases by roughly a factor two from 
$\eta =0$ to $\eta = 5$ in central PbPb collisions at the LHC (see Fig. 1 of \cite{ALICE:2016fbt}). 
Assuming $\pi^0$s have the same kinematic distribution as $\pi^{\pm}$s we can estimate 
$dN^{\pi^0}/dp_T\, dy \simeq 500/{\rm GeV}$ for $p_T < 500$ MeV and $4 < \eta < 5$.
This amounts to five $\pi^0$s, or equivalently ten background photons per event and per 10 MeV-bin of $p_T$ at forward rapidity to be compared with the number of photons in Fig.~\ref{fig:photon-number}. 
In other words, the signal-to-background ratio for the distributions in the upper panel of Fig.~\ref{fig:photon-number} is approximately  
\be
0.1 \lesssim \frac{S}{B} \lesssim 1 ~\mbox{ for }~ p_T  \lesssim 50 \,\mbox{MeV}~ \mbox{ and } ~ \eta \gtrsim 3 \, .
\ee
However, phenomenological models~\cite{Mehtar-Tani:2008hpn,Mehtar-Tani:2009wji} indicate that heavy-ion collisions are rather transparent to electric charge. A much better discrimination of bremsstrahlung signal to background is then required to establish bremsstrahlung photons. As seen from the left lower plot of Fig.~\ref{fig:photon-number},
 the experimental challenge of measuring bremsstrahlung for the stopping scenarios of ~\cite{Mehtar-Tani:2008hpn,Mehtar-Tani:2009wji} amounts to identifying as little as 
0.5 photons per event in the kinematic range $10\, {\rm MeV} < p_T < 20\, {\rm MeV}$, $3 \leq \eta \leq 4$. This is experimentally challenging.\footnote{There are interesting
parametric differences that may help to discriminate between bremsstrahlung 
photons and other photon sources. In particular, 
\begin{itemize}
\item{
$p_T$ dependence: 
$
\frac{dN^{\rm bgd}}{dp_T} \simeq \mbox{const.}   \quad \mbox{vs}  \quad \frac{dN^{\rm brems}}{dp_T} \propto \frac{1}{p_T}
$
}
\item{
centrality dependence:
$
\frac{d^2N^{\rm bgd}}{dp_T d\eta}  \propto N_{\rm part}  \quad \mbox{vs}  \quad \frac{d^2N^{\rm brems}}{dp_T d\eta} \propto Z^2  \propto N^2_{\rm part} 
$
}
\end{itemize}}
It therefore remains to be studied to what extent the experimental capabilities of future detectors like ALICE-3 can provide firm evidence for photon bremsstrahlung or whether they can only put experimental upper bounds on the bremsstrahlung spectrum that may allow one to disfavour some stopping scenarios. What our calculations clearly identify is the physics interest in equipping the kinematic range of ultra-soft photons of $\mathcal{O}(p_T \sim 10\, \text{ MeV})$ at forward rapidity with sensitive detectors. Our work is one of several recent works~\cite{Lebiedowicz:2021byo,Floerchinger:2021xhb} that point to the interest in more precise soft photon measurements at the LHC.

{\bf Acknowledgements}:  I thank Urs Wiedemann for collaboration and for a critical reading of this manuscript. I also 
thank Soyeon Cho, DongJo Kim, Min-Jung Kweon, J{\"u}rgen Schukraft, Martin V{\"o}lkl for useful questions and discussions during QM 2022 and in preparation of this manuscript.



\end{document}